\documentclass[cits]{PoS}

\def\Planck{{\sc Planck}}
\def\Point{\mbox{$\mathbf{P}$}}
\def\SpaceCraft{\mbox{$\mathbf{R}_{\mathrm{P}}$}}
\def\lsim{\,\lower2truept\hbox{${<\atop\hbox{\raise4truept\hbox{$\sim$}}}$}\,}
\def\gsim{\,\lower2truept\hbox{${>\atop\hbox{\raise4truept\hbox{$\sim$}}}$}\,}
\def\Ef{\mbox{$E_f$}}
\title{Simulating the Zody Emission in the Planck Mission}

\ShortTitle{Simulating the ZLE in the Planck Mission}

\author{\speaker{Michele Maris}%
         \\
        INAF-Osservatorio Astronomico di Trieste \\
       Via G.B.Tiepolo 11, I-34131 Trieste, Italy\\
        E-mail: \email{maris@oats.inaf.it}}

\author{ Carlo Burigana\\
        INAF-IASF Bologna\\
        Via Gobetti 101, I-40129 Bologna, Italy\\
        E-mail: \email{burigana@iasfbo.inaf.it}}

\author{Sandro Fogliani \\
        INAF-Osservatorio Astronomico di Trieste \\
       Via G.B.Tiepolo 11, I-34131 Trieste, Italy\\
        E-mail: \email{fogliani@oats.inaf.it}}

\abstract{
The increasing sensitivity of cosmic microwave background (CMB) missions will require 
to significantly improve the accuracy in the 
subtraction of the various sources of Galactic 
foreground, from the most relevant components
(synchrotron, dust and free-free emission) to those
usually considered of minor relevance in CMB experiments.
With respect to other Galactic diffuse components,
the Zodiacal Light Emission (ZLE) is peculiar, 
depending not only on the observing direction but also on the 
location of the observer within 
the Solar System: ZLE behaves then as a large scale, time-dependent foreground.
Starting from the existing far-infrared ZLE models,
we discuss the impact of ZLE contribution in CMB maps
and the level of contamination in time ordered data and maps
expected from the forthcoming \Planck\ space mission 
as well as the \Planck\ capability to increase our
knowledge of the ZLE properties.
}

\FullConference{CMB and Physics of the Early Universe\\
       20-22 April 2006\\
       Ischia, Italy}

\begin{document}

\section{Introduction}

The increasing sensitivity of cosmic microwave background (CMB) missions will require a careful 
subtraction of any source of weak ``Galactic'' 
foreground, asking to consider in 
CMB experiments 
other sources of Galactic foregrounds over the most relevant ones, i.e. 
synchrotron, dust, and free-free emission.
In this context, the \emph{Zodiacal Light Emission} (ZLE) due to the 
thermal emission from the cloud of Interplanetary Dust Particles (IDPs) 
permeating the Solar System represents a significant component.
At frequencies $\approx 10$~THz the ZLE dominates the sky emissivity,
and in preparing templates of Galactic emission from observations at 
these frequencies the contamination of ZLE has to be accurately 
accounted for, since 
%in an ideal experiment 
the ability to model and 
remove the ZLE will largely fix the final accuracy of the Galactic templates.
At frequencies below $\approx1$~THz the ZLE is subdominant compared to 
the Galaxy, but its surface brightness is still significant particularly 
in regions where the Galaxy emission is weak 
\cite{Fixsen:Dwek:2002,Maris:etal:2006}. 
It has also to be considered that the ZLE does not depend only on the
instrument pointing direction, \Point, but also on the position of the 
observer, \SpaceCraft, within the Solar System. Then, the ZLE behaves
as a time-dependent foreground and, when not properly removed, introduces
subtle systematic effects.

\section{Model and extrapolations}

As for the Galactic dust contamination, the modelling of the ZLE has to 
be based on far-infrared observations, at least in what regard the 
geometrical aspects of the IDPs distribution. Key data for the ZLE 
below 300~$\mu$m have been obtained by IRAS \cite{Wheelock:etal:1994}, 
COBE
\cite{Fixsen:Dwek:2002,Kelsall:etal:1998} and ISO
\cite{Reach:Abergel:Boulanger:1996,Reach:etal:2003}.
The starting point for this analysis is the COBE/DIRBE model for the 
ZLE which describes the expected local 3D emissivity within the IDPs
complex \cite{Kelsall:etal:1998}. 
In particular, among the various components in which 
the IDPs are distributed we refer here to the dominant \emph{Smooth}
component which accounts for more than 90\% of the ZLE. 
Then, for frequencies $f \lsim 1$~THz the total brightness of the ZLE integrated
along a given line-of-sight is

\begin{equation}\label{eq:ZOD}
  I_{f}(\Point,\SpaceCraft) = 
                    E_{f} \, Z_f{(\Point,\SpaceCraft)} \, ,
\end{equation}

 \noindent
where $Z_f{(\Point,\SpaceCraft)}$ gives the spatial dependence and
the \emph{Emissivity Factor} \Ef\ is a correction with respect to 
a pure blackbody emission law, related to
the composition and size distribution of dust grains.
Following 
\cite{Kelsall:etal:1998} it is assumed $\Ef = 1$ for $f=12$~THz.
The spatial dependence is given by the integral along the line-of-sight

\begin{equation}\label{eq:b:integrall}
                     Z_f{(\Point,\SpaceCraft)} = \int_{0}^{+\infty} ds
                    \;
                        \, N(\SpaceCraft+s\Point )
                        \, B_{f}\left(T(\SpaceCraft+s\Point )\right),
\end{equation}

 \noindent
where $s$ is the distance from the observer along \Point, 
$B_{f}(T)$ the 
blackbody brightness, $N(\mathbf{r})$ the dust density
at a given location in the Solar System 
(assuming $N(\mathbf{r}) \equiv 0$ for $|\mathbf{r}| > 5.2$~AU),
$T(\mathbf{r})$ the local dust temperature, assumed to scale as 
$|\mathbf{r}|^{-0.467}$.

In long duration CMB experiments the sky is observed 
%for a set of
%pointing directions 
while the observer is orbiting around the Sun and consequently
the ZLE will show seasonal modulations
at the level of $5 - 15\%$ \cite{Maris:etal:2006}.
Changes in the scanning strategy or observation epoch will result
in different realizations of ZLE sky maps even when the same set of
pointing directions is taken. To condensate into a statical map this 
dynamical information we exploit the cylindrical symmetry of COBE 
distribution of the IDPs Smooth component.
We develop a serie expansion of $Z_f{(\Point,\SpaceCraft)}$ about
an averaged orbit in the IDPs cloud reference frame 
\cite{Maris:etal:2006}. This is equivalent to calculate a kernel map 
for $Z_f{(\Point,\SpaceCraft)}$ which is good for a given ``nominal'' 
mission together with coefficients to be 
used in computing variations of the map for a range of possible 
variations in the mission.
This method allows at the same time the generation of data streams of 
ZLE signals for a given list of spacecraft positions and pointing 
directions, and the generation of specialized time averaged maps 
(e.g. yearly averaged maps) for a specific mission, orbit and scanning 
strategy~\footnote{Tables of serie expansions of ZLE for all of the \Planck\ 
frequencies, the related software in IDL, and documentation
may be required to the authors.}. 

The extrapolation of $E_f$ at frequencies below $1$~THz is a more 
delicate problem. COBE/DIRBE measures extend down to $f = 1.2$~THz,
which fixes the lowest frequency for which the COBE model provides
values of $E_f$ \cite{Kelsall:etal:1998}. However COBE/FIRAS provides
measures of ZLE averaged over the sky and over one year down-to 
$f \approx 3\times10^2$~GHz \cite{Fixsen:Dwek:2002} but with 
a not high S/N ratio. We then compare the COBE/DIRBE measures with
simulated yearly and full-sky averaged values of  
$Z_f{(\Point,\SpaceCraft)}$ in order to evaluate 
$E_f$. In this way we can obtain numerical estimates for 
these parameters 
%with 
%accuracies of about $\gsim20\%$ 
\cite{Maris:etal:2006}. 
Similar results are also obtained directly 
extrapolating $E_f$ values derived by COBE/DIRBE at $f < 12$~THz
down to the required frequencies. 
Tab.~\ref{tab:one} summarizes the results of these calculations giving the minimal
and maximal yearly averaged levels of contamination at 
frequencies relevant for the \Planck\ mission.

\begin{figure}
 \centering
 \begin{minipage}[t]{0.45\textwidth}
 \includegraphics[angle=90,width=1\textwidth]{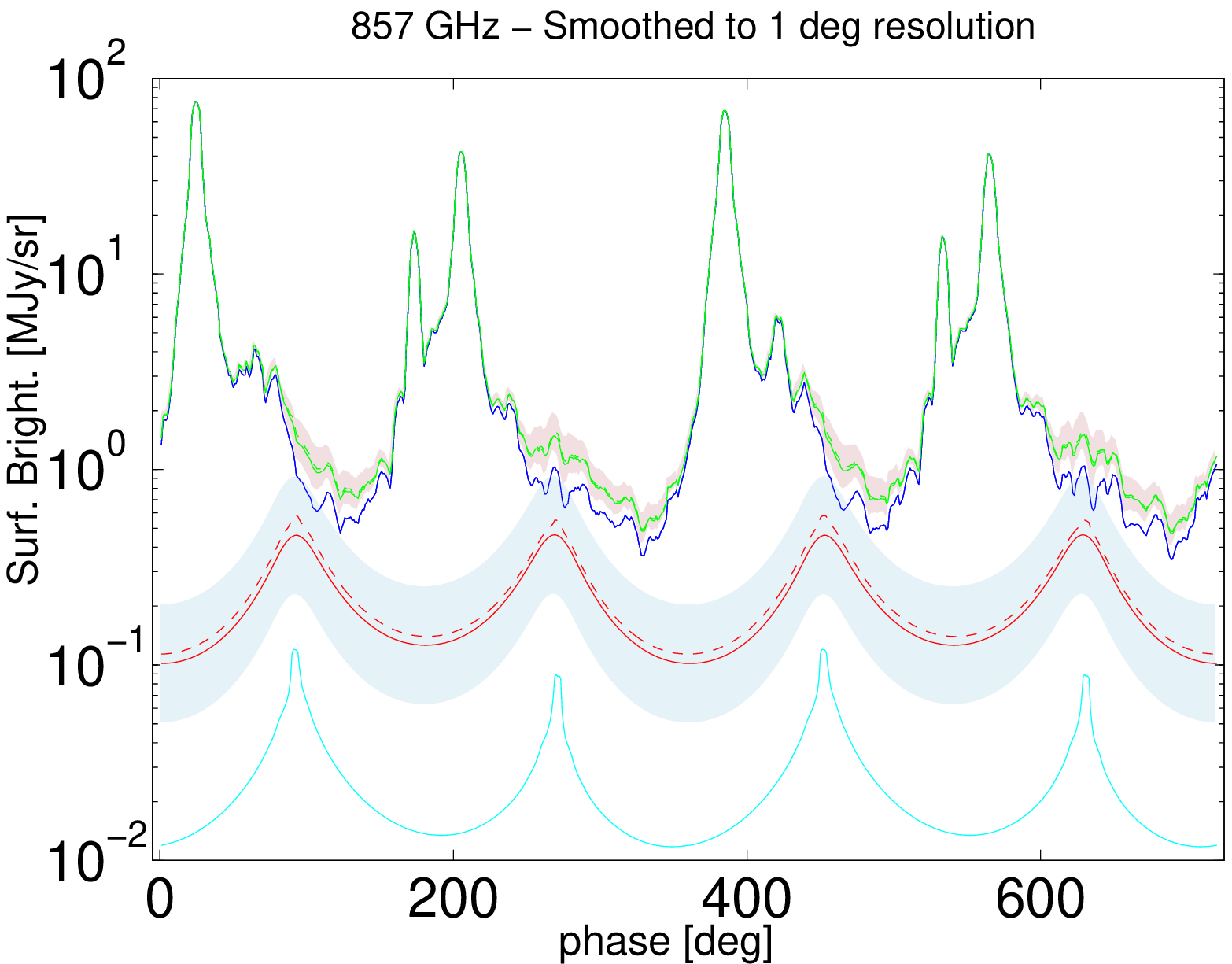}
  \end{minipage}
  \begin{minipage}[t]{0.45\textwidth}
  \includegraphics[angle=90,width=1.07\textwidth]{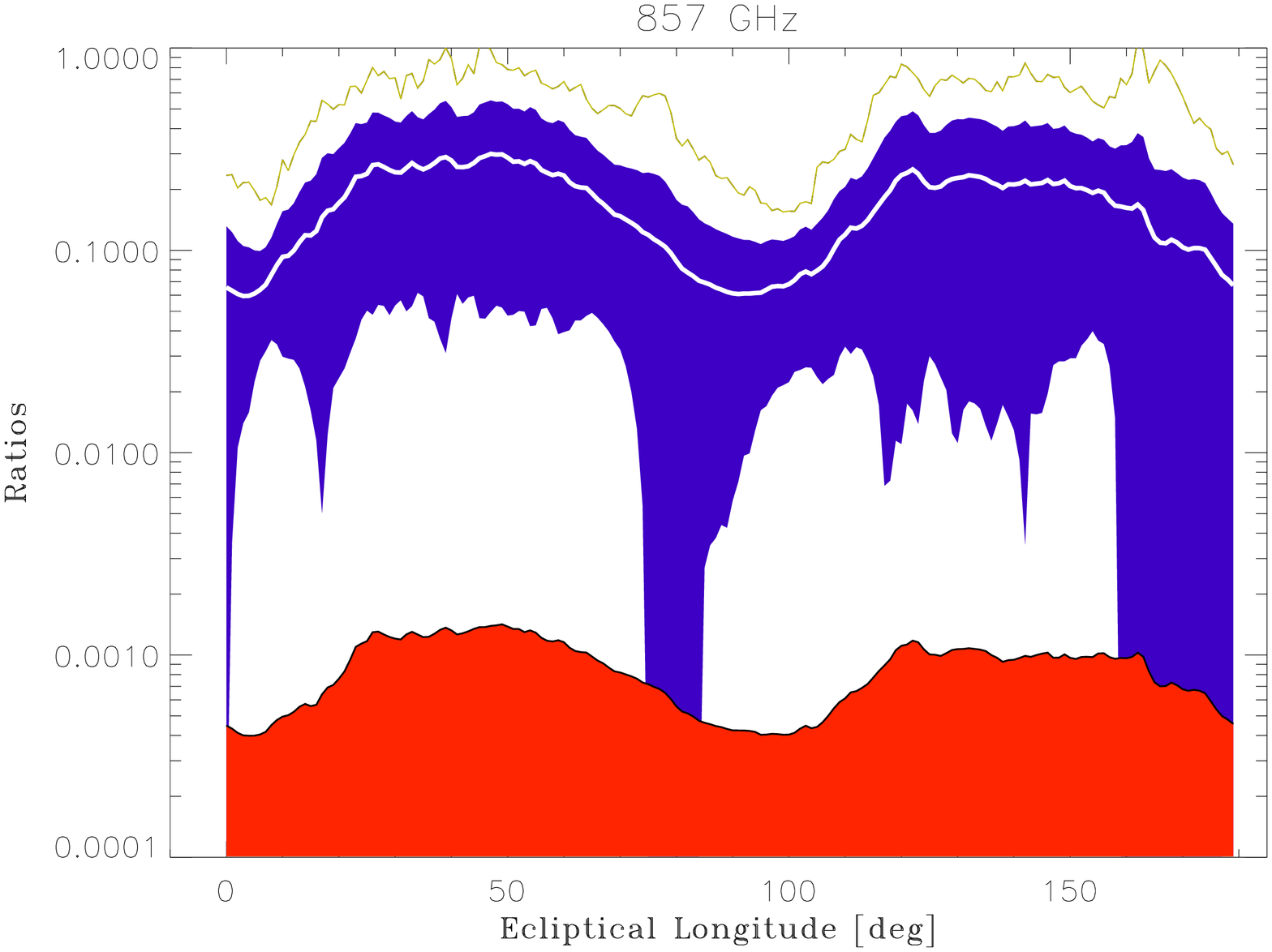}
  \end{minipage}
%  \rotatebox{0}
%{
% \includegraphics[width=0.45\textwidth]{maris_burigana_fig1_a.pdf}
% }
% &
%  \rotatebox{90}
%{
%  \includegraphics[width=0.35\textwidth]{maris_burigana_fig1_b.pdf}
% }
%\end{tabular}
\caption{
Left panel: simulated data stream of surface brightnesses (MJy/sr) 
expected at 857~GHz for the ZLE - smooth component (red), 
secondary components (cyan) and uncertainties in the prediction (bands) - 
the Galaxy  (green), and the sum of ZLE and Galaxy (blue). 
Also signals from ZLE
(red-dashed) as after summation of the Galaxy (green dashed)
are shown.
The ordinate is the phase along the scan circle.
%Bands are the uncertainties in the prediction.
 %
Right panel: relative averaged contribution of ZLE, and noise compared 
to the averaged Galaxy and computed over circular bands, as a function
of the ecliptical longitude of the spin axis vector.
}\label{fig:one}
 \end{figure}

The left frame of Fig.~\ref{fig:one} represents a typical realization 
of a ZLE signal at 857~GHz
(together with secondary components and uncertainties)
compared to the Galaxy.
The right frame shows the relative contribution of ZLE brighness and
instrumental noise respect to the Galaxy.
In the plot the ZLE and the noise contributions are averaged over a 
circular band of about $85^\circ$ of radius and drawn around 
an axis of given ecliptical longitude. The same is done for the Galaxy.
The red band is the noise, the white line the expected ZLE over
Galaxy averaged ratio, the blue band the $\pm1\sigma$ band, the 
yellow line the maximum ratio. The plotted values are ordered
as function of the ecliptical longitude of spin axis vector.
The data are calculated for patches of $1^\circ$ in radius.
The noise is referred to a 14 month mission (2 sky surveys).

%%%%%%%%%%%%%%%%%5
% FIG 1
%%%%%%%%%%%%%%%%%5

%%%%%%%%%%%%%%%%%5
% TAB 1
\newcommand{\EN}[2]{#1\times10^{#2}}
 \begin{table}
 \begin{center}
 \begin{tabular}{rcccc}
 \hline
 \hline
  &\multicolumn{4}{c}{Frequency Channell}\\
                      & 217 GHz      & 353 GHz & 545 GHz & 857 GHz \\
 \hline
 $E_f\;\;\;\;$          & $\EN{4}{-2}$ & $\EN{1}{-1}$ & $\EN{3}{-1}$& $\EN{6}{-1}$ \\
min  $I_{f}$   [MJy/sr] & $\EN{5}{-4}$ & $\EN{4}{-3}$ & $\EN{2}{-2}$ & $\EN{1}{-1}$ \\
mean  $I_{f}$  [MJy/sr] & $\EN{1}{-3}$ & $\EN{8}{-3}$ & $\EN{5}{-2}$ & $\EN{3}{-1}$\\
max  $I_{f}$   [MJy/sr] & $\EN{2}{-3}$ & $\EN{2}{-2}$ & $\EN{1}{-1}$
& $\EN{6}{-1}$ \\
 \hline
 \hline
 \end{tabular}
 \end{center}
 \caption{
Predicted values of $E_f$ and of the ZLE at the highest \Planck\ 
frequencies. The table reports minimum, mean and maximum ZLE
within a $\approx 50\%$ uncertainty for a one year mission.
 }\label{tab:one}
 \end{table}
%%%%%%%%%%%%%%%%%5

\section{Detection and removal of the ZLE}

It is evident that current information does not allow to reach 
an accuracy 
better then $\approx20\%$
in the ZLE removal;
this is due in particular to the uncertainties on \Ef\ below 1~THz. 
New, direct measures are necessary to improve the situation.
The ability of \Planck\ to 
detect the ZLE signal improving the accuracy in \Ef\ determination
has been analyzed in \cite{Maris:etal:2006} assuming that
the COBE model properly represents the 
spatial distribution of the ZLE leaving as a free parameter 
\Ef.
Two methods have been considered, the first based on the comparison
of a template map for the Galactic emission with a spatial template
for the ZLE calculated for given mission orbit and scanning strategy,
the second method is based on the 
comparison of observations of the same regions of sky 
taken at different epochs 
then exploiting the seasonal dependence in the ZLE observation.
In both cases, since the ZLE varies over scales of 
$\approx 10^\circ$,
one can consider template maps and observations 
at resolutions of
$\approx 1^\circ - 2^\circ$ (this alleviate also the contribution of
local features, as weak point sources or regions with peculiar frequency dependencies).
In this kinds of analysis,
it is important to apply cuts to the data excluding
regions where the Galaxy is very bright.
Simulations shows 
a typical absolute RMS uncertainty on \Ef\ determination
induced by the limited instrumental
sensitivity of $\sim 10^{-3}$,
% (1$\sigma$), 
$2.1 \times \sim 10^{-3}$ and $2.6 \times \sim 10^{-3}$
at 857~GHz, 545~GHz and 353~GHz, respectively.
For typical expected values of \Ef\
($\approx 0.65$, 0.26, 0.11 for 857~GHz, 545~GHz, 353~GHz)
 the \Planck\ sensitivity
will allow an \Ef\ recovery at
0.15\%, 0.8\% and 2.4\%
(1$\sigma$) accuracy at 857~GHz, 545~GHz and 353~GHz, respectively.
Of the most relevant systematic effects, 
pointing and sampling uncertainty, aberration of light, Doppler shift 
and relative calibration uncertainty, only the last one is found to be
really
%potentially the most 
critical,
% one, 
while the errors introduced by the other ones
%being
%expected 
are found to be significantly below the noise.
Then, relative calibration could ultimately determine 
the final accuracy in the
ZLE extraction from \Planck\ data.
For a relative calibration RMS error of $\sim 1\%$ $(0.1\%)$ on patches 
of $2^\circ$ radius, we find an absolute RMS error on \Ef\ of 
$\sim 0.01 - 0.04$ $(\sim 0.001 - 0.004)$ with only a weak dependence 
on the frequency in the range $\sim 300 - 900$~GHz, corresponding to 
relative errors on \Ef\ $\sim 4\%$, $10\%$, $23\%$
($\sim 0.4\%$, $1\%$, $2\%$), respectively at
857~GHz, 545~GHz, 353~GHz for the most likely \Ef\ values
expected on the basis of COBE/FIRAS data.
 %

%\acknowledgments

\end{document}